\documentclass[a4paper,12pt]{article}

\makeatletter
 
  \@addtoreset{equation}{section}
 \makeatother
\usepackage{amsfonts}
\usepackage{amssymb}
\usepackage{mathrsfs}
\usepackage{amsmath,amsthm}
\usepackage{amsmath}

\usepackage{graphicx}
\usepackage{color}
\usepackage{indentfirst}

\begin{document}
%nakamacro.tex(H120522;0730)
%\documentstyle[11pt]{article}
%\setlength{\textwidth}{6.5in}
%\setlength{\oddsidemargin}{0in}
%\setlength{\topmargin}{-0.52in}
%\setlength{\textheight}{9.0in}
%\setlength{\footskip}{0.7in}

%\newtheorem{definition}{Definition}
%\newtheorem{assumption}{$[$ A}
%\newtheorem{condition}{$[$ C}
%\newtheorem{lemma}{Lemma}
%\newtheorem{proposition}{Proposition}
%\newtheorem{theorem}{Theorem}
%\newtheorem{remark}{Remark}
%\newtheorem{example}{Example}
%--------------------------------------------------------------------------
%BOLD FACES
\def\bn{{\bf n}}
\def\A{{\bf A}}
\def\B{{\bf B}}
\def\C{{\bf C}}
\def\D{{\bf D}}
\def\E{{\bf E}}
\def\F{{\bf F}}
\def\G{{\bf G}}
\def\H{{\bf H}}
\def\I{{\bf I}}
\def\J{{\bf J}}
\def\K{{\bf K}}
\def\L{{\bf L}}
\def\M{{\bf M}}
\def\N{{\bf N}}
\def\O{{\bf O}}
\def\P{{\bf P}}
\def\Q{{\bf Q}}
\def\R{{\bf R}}
\def\S{{\bf S}}
\def\T{{\bf T}}
\def\U{{\bf U}}
\def\V{{\bf V}}
\def\W{{\bf W}}
\def\X{{\bf X}}
\def\Y{{\bf Y}}
\def\Z{{\bf Z}}
\def\cala{{\cal A}}
\def\calb{{\cal B}}
\def\calc{{\cal C}}
\def\cald{{\cal D}}
\def\cale{{\cal E}}
\def\calf{{\cal F}}
\def\calg{{\cal G}}
\def\calh{{\cal H}}
\def\cali{{\cal I}}
\def\calj{{\cal J}}
\def\calk{{\cal K}}
\def\call{{\cal L}}
\def\calm{{\cal M}}
\def\caln{{\cal N}}
\def\calo{{\cal O}}
\def\calp{{\cal P}}
\def\calq{{\cal Q}}
\def\calr{{\cal R}}
\def\cals{{\cal S}}
\def\calt{{\cal T}}
\def\calu{{\cal U}}
\def\calv{{\cal V}}
\def\calw{{\cal W}}
\def\calx{{\cal X}}
\def\caly{{\cal Y}}
\def\calz{{\cal Z}}
%
%YOKUTUKAUMONO
\def\sskip{\hspace{0.5cm}}
\def\simleq{ \raisebox{-.7ex}{\em $\stackrel{{\textstyle <}}{\sim}$} }
\def\leqsim{ \raisebox{-.7ex}{\em $\stackrel{{\textstyle <}}{\sim}$} }
\def\ep{\epsilon}
\def\half{\frac{1}{2}}
\def\iku{\rightarrow}
\def\Iku{\Rightarrow}
\def\ikup{\rightarrow^{p}}
\def\inclusion{\hookrightarrow}
\def\cadlag{c\`adl\`ag\ }
\def\up{\uparrow}
\def\down{\downarrow}
\def\doti{\Leftrightarrow}
\def\douti{\Leftrightarrow}
\def\dochi{\Leftrightarrow}
\def\douchi{\Leftrightarrow}%
%KAIGYOU,ARRAY
\def\yy{\\ && \nonumber \\}
\def\y{\vspace*{3mm}\\}
\def\nn{\nonumber}
\def\be{\begin{equation}}
\def\ee{\end{equation}}
\def\bea{\begin{eqnarray}}
\def\eea{\end{eqnarray}}
\def\beas{\begin{eqnarray*}}
\def\eeas{\end{eqnarray*}}
%
%KONO RONBUN DE TUKAU MONO
\def\hd{\hat{D}}
\def\hv{\hat{V}}
\def\hsd{{\hat{d}}}
\def\hx{\hat{X}}
\def\hsx{\hat{x}}
\def\bsx{\bar{x}}
\def\bsd{{\bar{d}}}
\def\bx{\bar{X}}
\def\ba{\bar{A}}
\def\bb{\bar{B}}
\def\bc{\bar{C}}
\def\bv{\bar{V}}
\def\balpha{\bar{\alpha}}
\def\bbalpha{\bar{\bar{\alpha}}}
\def\combi{\l(\begin{array}{c}\alpha\\ \beta \end{array}\r)}
\def\f{^{(1)}}
\def\s{^{(2)}}
\def\ss{^{(2)*}}
\def\l{\left}
\def\r{\right}
\def\a{\alpha}
\def\b{\beta}
\def\L{\Lambda}

\newtheorem{thm}{Theorem}
\newtheorem{lemma}{Lemma}
\newtheorem{prop}{Proposition}
\newtheorem{defn}{Definition}
\newtheorem{rem}{Remark}
\newtheorem{step}{Step}
\newtheorem{cor}{Corollary}

\newcommand{\Cov}{\mathop {\rm Cov}}
\newcommand{\Var}{\mathop {\rm Var}}
\renewcommand{\E}{\mathop {\rm E}}
\newcommand{\const }{\mathop {\rm const }}
\everymath {\displaystyle}

\newcommand{\ruby}[2]{
\leavevmode
\setbox0=\hbox{#1}
\setbox1=\hbox{\tiny #2}
\ifdim\wd0>\wd1 \dimen0=\wd0 \else \dimen0=\wd1 \fi
\hbox{
\kanjiskip=0pt plus 2fil
\xkanjiskip=0pt plus 2fil
\vbox{
\hbox to \dimen0{
\small \hfil#2\hfil}
\nointerlineskip
\hbox to \dimen0{\mathstrut\hfil#1\hfil}}}}

\def\qedsymbol{$\blacksquare$}
\renewcommand{\thefootnote }{\fnsymbol{footnote}}
\renewcommand{\refname }{References}

\everymath {\displaystyle}

\title{An Asymptotic Expansion Formula for Up-and-Out Barrier Option Price under Stochastic Volatility Model}
%\date{February 29, 2012}
\date{December 31, 2012}
\author{Takashi Kato\footnote{
Osaka University,}
\and Akihiko Takahashi\footnote{
The University of Tokyo,
}
\and 
Toshihiro Yamada\footnote{
The University of Tokyo \& MTEC} 
}
\maketitle

\allowdisplaybreaks

\abstract
This 
%In this 
paper derives a new semi closed-form approximation formula for pricing 
an up-and-out barrier option under a certain type of stochastic volatility model including SABR model 
by applying a rigorous asymptotic expansion method developed by Kato, Takahashi and Yamada \cite {1}. 
We also demonstrate the validity of our approximation method through
%show its  
numerical examples. 
%and justify the validity of our approximation  method.

\section{Introduction}
Numerical computation schemes for pricing barrier options 
have been a topic of great interest in mathematical finance and stochastic analysis. 
One of  the tractable approaches for evaluation of barrier options 
%calculating barrier options price 
is to derive an analytical approximation.
However, from the mathematical viewpoint, 
deriving an approximation formula by applying stochastic analysis 
is not an easy task since the Malliavin calculus approach as in Takahashi and Yamada \cite{3} cannot be directly applied. 
Recently, Kato, Takahashi and Yamada \cite {1} has provided a new asymptotic expansion method for 
the Cauchy--Dirichlet problem by developing  
%using 
a rigorous perturbation scheme in a partial differential equation (PDE), 
and as an example, derived an approximation formula for a down-and-out call option price 
under a stochastic volatility model.
% as an example. 
In this paper, 
we give a new asymptotic expansion formula for an up-and-out call option price 
under a stochastic volatility model 
which is widely used in trading practice. 
Moreover, we show 
%verify 
the validity of our formula through numerical experiments. 

\section{Asymptotic expansion formula for up-and-out barrier option prices}
Consider the following stochastic differential equation (SDE) in a stochastic volatility model:% for the currency option: 
\begin{eqnarray*}\label{SVdrift}
dS_{t}^{\varepsilon}&=&(c-q) S_{t}^{\varepsilon} dt+\sigma_t^{\varepsilon} S_{t}^{\varepsilon} dB_{t}^1, \\ S_0^{\varepsilon}&=&S,\\
d\sigma_{t}^{\varepsilon}&=&\varepsilon \lambda (\theta-\sigma_t^{\varepsilon})dt\\
&&+\varepsilon \nu \sigma_{t}^{\varepsilon}(\rho dB_{t}^1+\sqrt{1-\rho^2} dB_{t}^2), 
\\ \sigma_{0}^{\varepsilon}&=&\sigma,\nn 
\end{eqnarray*}
where $S, \sigma , c, q >0$, $\varepsilon \in [0,1)$, $\lambda, \theta, \nu>0$, $\rho \in [-1,1]$ and 
$B=(B^1,B^2)$ is a two dimensional standard Brownian motion. 
This model is motivated by pricing currency options. In this case, 
$c$ and $q$ represent a domestic interest rate and a foreign interest rate, respectively. 
The process $S^\varepsilon$ denotes a price of the underlying currency. % at time $t$. 
Our purpose is to evaluate
%calculate the price  of 
an {\it up-and-out} barrier option
with time-to-maturity $T-t$ and the upper barrier price $H (>S)$,
%the corresponding barrier option 
%of which 
and its initial value is represented
%given 
under a risk-neutral probability measure as follows:
%as in by the following: 
%, such as 
\begin{eqnarray*}
&&C_\mathrm {Barrier}^{SV,\varepsilon}(T-t, S)\\
&&\ \ \ \ \ \ \ = \E \left[e^{-c(T-t)} {f}(S^{\varepsilon } _{T-t} ) 1_{\{\tau _{(0,H)}(S^{\varepsilon }) > T-t\}}\right],
\end{eqnarray*}
where $f$ stands for a call option 
payoff function $f(s) = \max \{ s - K, 0 \}$ for some $K > 0$.
%, and $H(> S)$ is the upper barrier price. 
Here, the stopping time $\tau _{(0, H)}(S^{\varepsilon })$ is defined as
\begin{eqnarray*}
\tau _{(0, H)}(S^{\varepsilon }) = \inf \{ t\in [0, T] ; S^{\varepsilon }_t\notin (0, H)  \}
\  \   (\inf \emptyset := \infty). 
\end{eqnarray*} 

Remark that $C_\mathrm {Barrier}^{SV,\varepsilon}(T-t, S)$ has no closed-form solution and 
therefore we have to rely on some numerical method such as the Monte--Carlo simulation in order to calculate $C_\mathrm {Barrier}^{SV,\varepsilon}(T-t, S)$. 
However, when $\varepsilon=0$, 
$C_\mathrm {Barrier}^{SV,0}(T-t, S)$ corresponds to 
the up-and-out barrier option price in the Black-Scholes model which is known to 
%can 
be solved explicitly. 
Then, for $\varepsilon  > 0$, %$C_\mathrm {Barrier}^{SV,\varepsilon}(T-t, S)$, 
we are able to derive a {\it semi closed-form expansion} around 
$C_\mathrm {Barrier}^{SV,0}(T-t, S)$ when $\varepsilon \downarrow 0$.  
This is our main result and hereafter we show our approximation method for $C_\mathrm {Barrier}^{SV,\varepsilon}(T-t, S)$. 

Clearly, applying It\^o's formula, we can derive the SDE of logarithmic process of $S^\varepsilon _t$ as 
\begin{eqnarray*}
dX_{t}^{\varepsilon}&=&(c-q -\frac{1}{2}(\sigma_t^{\varepsilon})^2) dt+\sigma_t^{\varepsilon} dB_{t}^1, \\ X_{0}^{\varepsilon}&=&x \ := \ \log S. %\\
\end{eqnarray*}
Then we can rewrite $C_\mathrm {Barrier}^{SV,\varepsilon}(T-t, S)$ as 
\begin{eqnarray*}
&&C_\mathrm {Barrier}^{SV,\varepsilon}(T-t, e^x)\\
&&\ \ \ \ \ \ \ = \E \left[e^{-c(T-t)} \bar{f}(X^{\varepsilon } _{T-t} ) 1_{\{\tau _{D}(X^{\varepsilon }) > T-t\}}\right], 
\end{eqnarray*}
where $\bar{f}(x)=\max \{ e^x - K, 0 \}$ and $D = (-\infty, \log H )$. 
Note that 
\begin{eqnarray*}
\tau _{D}(X^{\varepsilon }) = \inf \{t\in [0, T] \ ; \ X^\varepsilon _t\notin D\} = \tau _{(0, H)}(S^{\varepsilon }). 
\end{eqnarray*}

Let $u^{\varepsilon}(t, x)=C_\mathrm {Barrier}^{SV,\varepsilon}(T-t,e^x)$ 
for $t\in [0, T]$ and $x\in {\bf R}$. 
Then $u^{\varepsilon}(t, x)$ satisfies the following PDE: 
\begin{eqnarray*}%\label{SV_PDE_drift}
\left\{ 
\begin{array}{ll}
\left(\frac{\partial}{\partial t}+\mathscr {L}^{\varepsilon}-c \right)u^{\varepsilon}(t,x) = 0, & (t, x)\in (0, T]\times D, \\
u^{\varepsilon}(T,x) = \bar{f}(x), & x \in \bar{D}, \\
u^{\varepsilon}(t,\log H) = 0, & t\in [0, T], 
\end{array}
\right. 
\end{eqnarray*}
where 
%Also, its generator is expressed as
\begin{eqnarray*}
\mathscr {L}^{\varepsilon}&=&\left(c-q -\frac{1}{2}\sigma^2 \right)\frac{\partial}{\partial x}+\frac{1}{2}\sigma^2\frac{\partial^2}{\partial x^2}\\
&&+\varepsilon \rho \nu\sigma^2 \frac{\partial^2}{\partial x \partial \sigma}+\varepsilon \lambda (\theta-\sigma)\frac{\partial}{\partial \sigma}
+\varepsilon^2 \frac{1}{2}\nu^2\sigma^2 \frac{\partial^2}{\partial  \sigma^2}.\label{Gene_ep}
\end{eqnarray*}
%Define $\tilde{\mathscr {L}}^0_k$ as 
%\begin{eqnarray}
%\tilde{\mathscr {L}}^0_k = 
%\frac{1}{k!} 
%\frac{\partial ^k}{\partial \varepsilon ^k}\mathscr {L}^{\varepsilon}|_{\varepsilon=0}
% . \label{g_expansion}
%\end{eqnarray}
%Remark that 
%$\tilde{\mathscr {L}}_1^0$  is given by 
%\begin{eqnarray}
%\tilde{\mathscr {L}}_1^0 &=&\rho \sigma^2{\partial^2\over \partial x \partial \sigma}+\lambda (\theta-\sigma)\frac{\partial}{\partial \sigma}. \label{Gene_1}
%\end{eqnarray}
%the $0$-th order $u^0$ as
%\begin{eqnarray*}
%u^{0}(t,x)&=&P^{D}_{T-t}{\bar f}(x)\\
%&=&\E[e^{-c(T-t)}{\bar f}(X_{T-t}^{x,0})1_{\{\tau _{D}(X^{0, x}) > T-t\}} ]. 
%\end{eqnarray*}
%Set $\alpha=c-q$ for brevity. 
As mentioned above, when $\varepsilon = 0$, we can obtain the explicit value of $u^0(t, x)$. 
In this case, $u^0(t, x) = %C^{BS}_\mathrm{Barrier}=
C^{BS}_\mathrm{Barrier}(T-t ,e^x,\sigma,H)$ 
represents the price of the up-and-out barrier call option under the Black--Scholes model. We have 
\begin{eqnarray*}
C_{\mathrm{Barrier}}^{BS}=C_{\mathrm{Vanilla}}^{BS}-C,
\end{eqnarray*}
where 
\begin{eqnarray*}
C_{\mathrm{Vanilla}}^{BS}&=&e^x e^{-qT}N(d_1)-Ke^{-cT}N(d_2),\\
C&=&e^x e^{-qT}N(x_1)-Ke^{-cT}N(x_2)\\
&&-e^x e^{-qT} \left( \frac{H}{e^x} \right)^{2\lambda}[ N(-y)-N(-y_1) ]\\
&&+ Ke^{-cT} \left( \frac{H}{e^x} \right)^{2\lambda-2}\\
&&\ \ \times [ N(-y+\sigma \sqrt{T})-N(-y_1+\sigma \sqrt{T}) ]
\end{eqnarray*}
with
\begin{eqnarray*}
x_1
&=&
\frac{x- \log H +(c-q)T+1/2 \sigma^2 {T}  }{\sigma \sqrt{T}},
\\
x_2&=&x_1-\sigma \sqrt{T},\\
\lambda&=&\frac{(c-q)}{\sigma^2}+ \frac{1}{2}, \\
y&=&
\frac{2 \log H -x-\log K +(c-q)T+1/2 \sigma^2 {T}  }{\sigma \sqrt{T}},\\
\\
y_1&=&
\frac{\log H- x +(c-q)T+1/2 \sigma^2 {T}  }{\sigma \sqrt{T}}.
\end{eqnarray*}
See Hull \cite{2} for the details.\\  

We can represent $u^0(t, x) = \bar{P}^D_t\bar{f}(x)$ by using a semi-group $(\bar{P}^D_t)_t$ defined as 
\begin{eqnarray}
{\bar P}_{s}^{D}g(x)&=&\int_{-\infty}^{\log H} e^{-c s} (1-e^{-\frac{2(\log H-x)(\log H-y)}{\sigma^2s}})\nonumber\\
&&\ \ \times \frac{1}{\sqrt{2 \pi \sigma^2 s}}
e^{-\frac{(y-x-(c-q-\frac{1}{2}\sigma^2) s )^2}{2\sigma^2 s}}
g(y) dy\nonumber\\ \label{semigroup}
\end{eqnarray}
for a continuous function $g$ with polynomial growth rate which satisfies $g(x)=0$ on $\partial D$. 

The main result of Kato, Takahashi and Yamada \cite {1} suggests the following approximation formula.\\
\\
{\bf [Asymptotic expansion formula]}
\begin{eqnarray*}
&&u^{\varepsilon}(t,x)= C_{\mathrm{Barrier}}^{BS}\\
&&\ \ \ \ +\varepsilon   e^{-c(T-t)}\int_{0}^{T-t}{\bar P}^{D}_{s} \tilde{\mathscr {L}}^0_{1} {\bar P}^{D}_{T-t-s} {\bar f}(x)ds+O(\varepsilon^2), 
\end{eqnarray*}
where 
\begin{eqnarray*}
\tilde{\mathscr {L}}^0_1 = 
\frac{\partial }{\partial \varepsilon }\mathscr {L}^{\varepsilon}|_{\varepsilon=0} = 
\rho \sigma^2{\partial^2\over \partial x \partial \sigma}+\lambda (\theta-\sigma)\frac{\partial}{\partial \sigma}. %\label{Gene_1}
\label{g_expansion}
\end{eqnarray*}
\noindent\\

Using (\ref{semigroup}), the term $\int_{0}^{T-t}{\bar P}^{D}_{s} \tilde{\mathscr {L}}^0_{1} {\bar P}^{D}_{T-t-s} {\bar f}(x)ds$ is expressed as follows:
\begin{eqnarray}
&&\int_{0}^{T-t}{\bar P}^{D}_{s} \tilde{\mathscr {L}}^0_{1} {\bar P}^{D}_{T-t-s} {\bar f}(x)ds\nonumber\\
&=&\int_{0}^{T-t} \int_{-\infty}^{\log H} e^{-c s} (1-e^{-\frac{2(\log H-x)(\log H-y)}{\sigma^2s}})\nonumber\\
&&\ \ \times \frac{1}{\sqrt{2 \pi \sigma^2 s}}
e^{-\frac{(y-x-(c-q-\frac{1}{2}\sigma^2) s )^2}{2\sigma^2 s}}
\tilde{\mathscr {L}}^0_{1} {\bar P}_{T-t-s}^{D}\bar{f}(y)dyds.\nonumber\\
\label{Approx_term}
\end{eqnarray}

%\noindent\\
We are able to compute %the approximation term as follows. 
the integrand of the right hand side of the above formula (\ref{Approx_term}) as 
\begin{eqnarray*}
&&\tilde{\mathscr {L}}^0_{1} {\bar P}_{T-t}^{D}\bar{f}(x)\\
&=&e^{c(T-t)} \Biggl\{ \rho \sigma^2 \frac{\partial^2}{\partial x \partial \sigma} C_{\mathrm{Barrier}}^{BS}(T-t,e^x,\sigma)\\
&&+\lambda (\theta-\sigma) \frac{\partial}{\partial \sigma} C_{\mathrm{Barrier}}^{BS}(T-t,e^x,\sigma) \Biggr\} .
\end{eqnarray*}
Here, $\frac{\partial}{\partial \sigma}C_{\mathrm{Barrier}}^{BS}(T,e^x)$ and $\frac{\partial^2}{\partial x \partial \sigma}C_{\mathrm{Barrier}}^{BS}(T,e^x)$ are concretely expressed as follows:
\begin{eqnarray*}
&&\frac{\partial}{\partial \sigma}C_{\mathrm{Barrier}}^{BS}(T,e^x)\\
&=&e^{-qT} e^x n(d_1) \sqrt{T}\\
&&-e^{-qT} e^x n(x_1) \sqrt{T}-(H-K)e^{-cT} n(x_2) \frac{-x_1}{\sigma}\\
&&+e^x e^{-qT} \left( \frac{H}{e^x} \right)^{2\lambda}\\
&&\times \Biggl\{ (\log H - x) \frac{-4(c-q)}{\sigma^3}[ N(-y)-N(-y_1) ]\\
&& + [n(y)\frac{y'}{\sigma} -n(y_1)\frac{y'_1}{\sigma}] \Biggr\}\\
&&-Ke^{-cT} \left( \frac{H}{e^x} \right)^{2\lambda-2} \\
&&\times \Biggl\{ (\log H - x) \frac{-4(c-q)}{\sigma^3}[ N(-y')-N(-y'_1) ]\\
&&+ [n(y')\frac{y}{\sigma}-n(y'_1)\frac{y_1}{\sigma}] \Biggr\}, 
\end{eqnarray*}
\begin{eqnarray*}
&&\frac{\partial^2}{\partial x \partial \sigma}C_{\mathrm{Barrier}}^{BS}(T,e^x)
=e^{-qT} e^{x}n(d_1)(-d_2)\frac{1}{\sigma}\\
&&-e^{-qT} e^{x}n(x_1)(-x_2)\frac{1}{\sigma}\\
&&-(H-K)e^{-cT} \frac{n(x_2)}{\sigma^2 \sqrt{T}} \{{x_1 x_2}-1 \}\\
&&+\frac{4(c-q)}{\sigma^3} \{ (-1+2\lambda )(\log H - x)+1 \} \\
&&\times e^x e^{-qT} \left( \frac{H}{e^x} \right)^{2\lambda} [ N(-y)-N(-y_1) ]
\\
&&+e^x e^{-qT} \left( \frac{H}{e^x} \right)^{2\lambda} [n(y)\frac{y'}{\sigma}-n(y_1)\frac{y'_1}{\sigma}]\\
&&\times \left(1 -2 \lambda \left( \frac{H}{e^x} \right)^{2\lambda}  \right)\\
&&-e^x e^{-qT} \left( \frac{H}{e^x} \right)^{2\lambda} (\log H - x)\\
&&\times  \frac{4(c-q)}{\sigma^3} \left( n(y)\left(\frac{1}{\sigma \sqrt{T}}\right)-n(y_1)\left(\frac{1}{\sigma \sqrt{T}}\right) \right)\\
&&+e^x e^{-qT} \left(  \frac{H}{e^x} \right)^{2\lambda}\\
&&\times  \left( n(y)\frac{1}{\sigma^2 \sqrt{T}}(yy'-1)-n(y_1)\frac{1}{\sigma^2 \sqrt{T}}(y_1y'_1-1)  \right)\\
&&-Ke^{-cT}[ N(y')-N(y'_1) ]\\
&&\times \left(  \left( \frac{H}{e^x} \right)^{2\lambda-2}\frac{4(c-q)}{\sigma^3} \{ (2 \lambda-2) (\log H - x)+1\}  \right)\\
&&+Ke^{-cT} \left( \frac{H}{e^x} \right)^{2\lambda-2} (\log H - x) \frac{4(c-q)}{\sigma^3}\\
&&\times  \left( n(y') \frac{1}{\sigma \sqrt{T}}-n(y'_1) \frac{1}{\sigma \sqrt{T}} \right)\\
&&+ Ke^{-cT}  (2 \lambda-2) \left( \frac{H}{e^x} \right)^{2\lambda-2}\\
&&\times \left( n(y') \frac{y}{\sigma}-n(y'_1) \frac{y_1}{\sigma}  \right)\\
&&-Ke^{-cT} \left( \frac{H}{e^x} \right)^{2\lambda-2}\\
&&\times  \left(n(y')\frac{1}{\sigma^2 \sqrt{T}}(y'y-1)-n(y'_1)\frac{1}{\sigma^2 \sqrt{T}}(y'_1 y_1-1)  \right) , 
\end{eqnarray*}
where
\begin{eqnarray*}
&&y'=\frac{2 \log H-x-\log K+(c-q)T-\frac{1}{2} \sigma^2 {T}  }{\sigma \sqrt{T}},\\
&&y'_1=\frac{\log H- x +(c-q)T-\frac{1}{2} \sigma^2 {T}  }{\sigma \sqrt{T}}.
\end{eqnarray*}

\section{Numerical Examples}
In this section we show numerical examples for pricing
%of 
European up-and-out barrier call options under SABR volatility model ($\lambda=0$)
as an illustrative purpose. 
By the asymptotic expansion formula in the previous section, we see 
\begin{eqnarray*}
C_\mathrm {Barrier}^{SV,\varepsilon}(T,S) &\simeq& C_\mathrm {Barrier}^{BS}(T,S)\\
&&+\varepsilon e^{-c T}\int_{0}^{T}{\bar P}^{D}_{s} \tilde{\mathscr {L}}^0_{1} {\bar P}^{D}_{T-s} {f}(S)ds.
\end{eqnarray*} 
Let us define {\bf AE first} and {\bf AE zeroth} as 
\begin{eqnarray*}
\mbox{{\bf AE first}}&=& C_\mathrm {Barrier}^{BS}(T,S)\\
&&+\varepsilon e^{-c T}\int_{0}^{T}{\bar P}^{D}_{s} \tilde{\mathscr {L}}^0_{1} {\bar P}^{D}_{T-s} {f}(S)ds,\\
\mbox{{\bf AE zeroth}}&=&C_\mathrm {Barrier}^{BS}(T,S).
\end{eqnarray*} 

Below we list the numerical examples, [Case 1] -- [Case 6],
%We list the numerical examples below, 
where
the numbers in the parentheses show the error rates (\%)
relative to the benchmark prices of $C_\mathrm {Barrier}^{SV,\varepsilon}(T,S)$ 
which are computed by Monte--Carlo simulations with $100,000$ time steps and $1,000,000$ trials 
(denoted by {\bf MC}).
We check the accuracy of our approximation formula
by changing the model parameters. 

Apparently, 
\vspace{1mm}
our approximation formula {\bf AE first} improves the accuracy for $C_\mathrm {Barrier}^{SV,\varepsilon}(T,S)$, 
and it is observed that the approximation term $\varepsilon e^{-c T}\int_{0}^{T}{\bar P}^{D}_{s} \tilde{\mathscr {L}}^0_{1} {\bar P}^{D}_{T-s} {f}(S)ds$ accurately compensates for the difference between 
$C_\mathrm {Barrier}^{SV,\varepsilon}(T,S)$ 
and $C_\mathrm {Barrier}^{BS}(T, S)$, which confirms the validity of our method. \\
%\noindent\\
\noindent\\
For all cases, we set
$S=100$, $\sigma=0.2$, $c=0.0$, $q=0.0$,
$\rho=-0.5$, $\varepsilon \lambda=0.0$,
$\theta=0.0$ and $T=1.0$.
In Case 1, 2 and 3, 
given $\varepsilon \nu=0.1$,
the upper bound price is set as
$H=120, 130, 140$, respectively,
while in Case 4, 5 and 6, 
given $\varepsilon \nu=0.2$,
%we set the upper bound price as
$H$ is set as
$120, 130, 140$, respectively. 
Particularly, for the case of $\varepsilon \nu=0.2$ (that is, 
higher volatility of volatility case,  Case 4, 5 and 6), 
we remark that the errors of the approximation become  slightly larger. 
However, as observed in comparison between 
{\bf AE first} and {\bf AE zeroth}, 
we are convinced that the higher order expansion
improves the approximation further,
which will be investigated in our next research.\\
%compensates the errors and this is a next research topic one of our next research topics. \\

\noindent
$[{\bf Case 1}]$\\
%The model parameters are given as follows. 
%where the upper bound price is changed from $H=120$, $H=130$, $H=140$.
\begin{eqnarray*}
&&
S=100,\ \sigma=0.2,\  c=0.0, \ q=0.0,\  \varepsilon \nu=0.1, \\
&&
\rho=-0.5,\ \varepsilon \lambda=0.0,\ \theta=0.0,\ H=120,\ T=1.0.
%&&
 %K=100,\ 102,\ 105.
\end{eqnarray*}
\begin{table}[ht]
\begin{center}
\caption{
%Calculated values of the 
up-and-out barrier option prices and the relative errors (Case 1)}%{\bf [Case 1]} Up-and-Out Barrier Option}
\label{fig3}
\begin{tabular}{|c|c|c|c| } \hline
{Strike: $K$} & \multicolumn{1}{|c|}{{\bf MC}} & \multicolumn{1}{|c}{\mbox{{\bf AE} {\bf first}}}  
& \multicolumn{1}{|c|}{\mbox{{\bf AE} {\bf zeroth}}} 
  \\\hline\hline
100  & 1.204 &  1.188 (-1.35\%) & 1.105  (-8.25\%) \\
102  & 0.882 & 0.869 (-1.44\%) &  0.804 (-8.78\%) \\
105 & 0.512 &  0.504 (-1.62\%) &  0.463 (-9.59\%) \\
\hline
\end{tabular}
\end{center}
\end{table}
\noindent\\ \\ \\ \\
$[{\bf Case 2}]$\\
%The model parameters are given as follows.
\begin{eqnarray*}
&&
S=100,\ \sigma=0.2,\  c=0.0, \ q=0.0,\  \varepsilon \nu=0.1, \\
&&
\rho=-0.5,\ \varepsilon \lambda=0.0,\ \theta=0.0,\ H=130,\ T=1.0.
%&&
 %K=100,\ 102,\ 105.
\end{eqnarray*}
\begin{table}[ht]
\begin{center}
\caption{
%Calculated values of the 
up-and-out barrier option prices and the relative errors (Case 2)}%{\bf [Case 1]} Up-and-Out Barrier Option}
\label{fig3}
\begin{tabular}{|c|c|c|c| } \hline
{Strike: $K$} & \multicolumn{1}{|c|}{{\bf MC}} & \multicolumn{1}{|c}{\mbox{{\bf AE} {\bf first}}}  
& \multicolumn{1}{|c|}{\mbox{{\bf AE} {\bf zeroth}}} 
  \\\hline\hline
100  & 3.216 & 3.200 (-0.49\%) & 2.966 (-7.78\%) \\
102  & 2.621 & 2.607 (-0.55\%) & 2.406 (-8.22\%) \\
105 & 1.869 & 1.857 (-0.69\%) & 1.702 (-8.93\%) \\
\hline
\end{tabular}
\end{center}
\end{table}
\noindent\\ \\ \\
$[{\bf Case 3}]$\\
%The model parameters are given as follows.
\begin{eqnarray*}
&&
S=100,\ \sigma=0.2,\  c=0.0, \ q=0.0, \varepsilon \nu=0.1,\\
&&
\rho=-0.5, \ 
\varepsilon \lambda=0.0,\ \theta=0.0,\
H=140,\ T=1.0.%\  K=100,\ 102,\ 105.
\end{eqnarray*}
\begin{table}[ht]
\begin{center}
\caption{
%Calculated values of the 
up-and-out barrier option prices and the relative errors (Case 3)}%{\bf [Case 2]} Up-and-Out Barrier Option}
\label{fig5}
\begin{tabular}{|c|c|c|c| } \hline
{Strike: $K$} & \multicolumn{1}{|c|}{{\bf MC}} & \multicolumn{1}{|c}{\mbox{{\bf AE} {\bf first}}}  
& \multicolumn{1}{|c|}{\mbox{{\bf AE} {\bf zeroth}}} 
  \\\hline\hline
100  & 5.184 & 5.186 (0.05\%) & 4.847 (-6.49\%) \\
102  & 4.420 & 4.423 (0.06\%) & 4.121 (-6.77\%) \\
105 & 3.420 & 3.422 (0.06\%) & 3.174 (-7.19\%) \\
\hline
\end{tabular}
\end{center}
\end{table}
\noindent\\ \\ \\
$[{\bf Case 4}]$\\
%The model parameters are given as follows.
% where the upper bound price is changed from $H=130$ to $H=140$.
\begin{eqnarray*}
&&
S=100,\ \sigma=0.2,\  c=0.0, \ q=0.0, \varepsilon \nu=0.2,\\
&&
\rho=-0.5, \ 
\varepsilon \lambda=0.0,\ \theta=0.0,\
H=120,\ T=1.0.%\  K=100,\ 102,\ 105.
\end{eqnarray*}
\begin{table}[ht]
\begin{center}
\caption{
%Calculated values of the 
up-and-out barrier option prices and the relative errors (Case 4)}%{\bf [Case 2]} Up-and-Out Barrier Option}
\label{fig5}
\begin{tabular}{|c|c|c|c| } \hline
{Strike: $K$} & \multicolumn{1}{|c|}{{\bf MC}} & \multicolumn{1}{|c}{\mbox{{\bf AE} {\bf first}}}  
& \multicolumn{1}{|c|}{\mbox{{\bf AE} {\bf zeroth}}} 
  \\\hline\hline
100  & 1.317 & 1.271 (-3.51\%) & 1.105 (-16.12\%) \\
102  & 0.971 & 0.934 (-3.83\%) & 0.804 (-17.15\%) \\
105 & 0.569 & 0.545 (-4.30\%) &  0.463 (-18.65\%) \\
\hline
\end{tabular}
\end{center}
\end{table}
\noindent\\ \\ \\
$[{\bf Case 5}]$\\
%The model parameters are given as follows.
% where the upper bound price is changed from $H=130$ to $H=140$.
\begin{eqnarray*}
&&
S=100,\ \sigma=0.2,\  c=0.0, \ q=0.0, \varepsilon \nu=0.2,\\
&&
\rho=-0.5, \ 
\varepsilon \lambda=0.0,\ \theta=0.0,\
H=130,\ T=1.0.%\  K=100,\ 102,\ 105.
\end{eqnarray*}
\begin{table}[ht]
\begin{center}
\caption{
%Calculated values of the 
up-and-out barrier option prices and the relative errors (Case 5)}%{\bf [Case 2]} Up-and-Out Barrier Option}
\label{fig5}
\begin{tabular}{|c|c|c|c| } \hline
{Strike: $K$} & \multicolumn{1}{|c|}{{\bf MC}} & \multicolumn{1}{|c}{\mbox{{\bf AE} {\bf first}}}  
& \multicolumn{1}{|c|}{\mbox{{\bf AE} {\bf zeroth}}} 
  \\\hline\hline
100  & 3.475  & 3.435 (-1.15\%) &  2.966 (-14.66\%) \\
102  & 2.844 & 2.808 (-1.27\%) &  2.406 (-15.42\%) \\
105 & 2.041 & 2.011 (-1.48\%) &   1.702 (-16.58\%) \\
\hline
\end{tabular}
\end{center}
\end{table}
\noindent\\ \\ \\
$[{\bf Case 6}]$\\
%The model parameters are given as follows.
% where the upper bound price is changed from $H=130$ to $H=140$.
\begin{eqnarray*}
&&
S=100,\ \sigma=0.2,\  c=0.0, \ q=0.0, \varepsilon \nu=0.2,\\
&&
\rho=-0.5, \ 
\varepsilon \lambda=0.0,\ \theta=0.0,\
H=140,\ T=1.0.%\  K=100,\ 102,\ 105.
\end{eqnarray*}
\begin{table}[ht]
\begin{center}
\caption{
%Calculated values of the 
up-and-out barrier option prices and the relative errors (Case 6)}%{\bf [Case 2]} Up-and-Out Barrier Option}
\label{fig5}
\begin{tabular}{|c|c|c|c| } \hline
{Strike: $K$} & \multicolumn{1}{|c|}{{\bf MC}} & \multicolumn{1}{|c}{\mbox{{\bf AE} {\bf first}}}  
& \multicolumn{1}{|c|}{\mbox{{\bf AE} {\bf zeroth}}} 
  \\\hline\hline
100  & 5.483 & 5.526 (0.78\%) &  4.847  (-11.59\%) \\
102  & 4.683 & 4.725 (0.85\%) &  4.121 (-12.03\%) \\
105 & 3.635 & 3.670 (0.97\%) &  3.174  (-12.68\%) \\
\hline
\end{tabular}
\end{center}
\end{table}


\begin{thebibliography}{99}
\bibitem{1}
T. Kato, A. Takahashi and T. Yamada, An asymptotic expansion for solutions of Cauchy-Dirichlet problem for second order parabolic PDEs and its application to pricing barrier options, {arXiv preprint}, (2012).
\bibitem {2}J.C. Hull,  
Options, Futures, and Other Derivatives, 6-th Edition,  
Prentice Hall, 2005
\bibitem{3}
A.Takahashi and T.Yamada, An asymptotic expansion with push-down of Malliavin weights, SIAM Journal on Financial Mathematics, {\bf 3}, (2012), 95--136.
\end{thebibliography}
\end{document}